# On the Role of Non-Periodic Orbits in the Semiclassical Quantization of the Truncated Hyperbola Billiard

R. Aurich[1], T. Hesse[2], and F. Steiner[1]

II. Institut für Theoretische Physik, Universität Hamburg,  
Luruper Chaussee 149, 22761 Hamburg, Germany

Based on an accurate computation of the first 1851 quantal energy levels of the truncated hyperbola billiard, we have found an anomalous long-range modulation in the integrated level density. It is shown that the observed anomaly can be explained by an additional term in Gutzwiller's trace formula. This term is given as a sum over families of closed, non-periodic orbits which are reflected in a point of the billiard boundary where the boundary is continuously differentiable, but its curvature radius changes discontinuously.

[1] Supported by Deutsche Forschungsgemeinschaft under Contract No. DFG–Ste 241/4–6 and 6–1  
[2] Supported by Konrad-Adenauer-Stiftung

During the last few years a lot of work has been devoted to the semiclassical quantization of classically chaotic Hamiltonian systems - especially Euclidean billiards and the geodesic flow on negatively curved Riemannian surfaces. The main effort of these examinations lay in probing Gutzwiller's trace formula, see eq. (8), that was first derived in [1]. For a classically chaotic billiard system with a smooth boundary and unstable, isolated periodic orbits only, this formula gives a semiclassical relation between geometrical properties of the periodic orbits and the quantum mechanical energy spectrum. For certain surfaces of constant negative curvature, this formula coincides with the exact, rigorously proven Selberg trace formula of analytic number theory [2].

Recently billiard systems have been examined which break the conditions stated above for applying Gutzwiller's trace formula. One example is the well-known Bunimovich stadium billiard [3]. In this system there exists a family of neutral periodic orbits (so-called bouncing ball orbits), and in addition the boundary curvature changes discontinuously in four points, which we will denote by *critical points*. In [4] and [5] the contribution of the bouncing ball orbits to the quantum mechanical level density has been calculated. Also, the effect of the boundary discontinuities has been studied in these papers. In the stadium billiard it turns out that *periodic orbits*, which are reflected in one of the critical points, need special consideration. In [4] and [5] the contribution of their single traversal has been calculated for the desymmetrized and the full stadium billiard respectively. In this Letter we derive the semiclassical contribution of *non-periodic* "critical orbits" to the trace formula and show that under certain conditions their contribution even exceeds the periodic-orbit contribution.

The investigated system is the truncated hyperbola billiard, see fig. 1, which is a variation of the unbounded hyperbola billiard studied previously in [6]. Its boundary consists of the $x$-axis, the diagonal $y = x$, and the hyperbola $y = \frac{1}{x}$, truncated at the point $A = \left(\alpha, \frac{1}{\alpha}\right)$ by a circular arc perpendicular to the $x$-axis. At the connection point $A$, the boundary is continuously differentiable, but its curvature radius changes discontinuously from $R^+ = -\frac{1}{2}\left(\alpha^2 + \frac{1}{\alpha^2}\right)^{\frac{3}{2}}$ (hyperbola) to $R^- = \left(\frac{1}{\alpha^2} + \frac{1}{\alpha^6}\right)^{\frac{1}{2}}$ (circular arc). In our numerical calculations presented below, the value of the parameter $\alpha$ is chosen to be $\alpha = 7.1175806$. The motivation for investigating this billiard (with the given value for $\alpha$) stems from a recent experiment [7] with a flat superconducting microwave cavity shaped like the truncated hyperbola billiard. For frequencies below approximately 20 GHz, only TM modes are excited which are described by the two-dimensional Helmholtz equation $-\Delta E_z = k^2 E_z$ with Dirichlet boundary conditions. This wave equation coincides with the two-dimensional Schrödinger equation $-\frac{\hbar^2}{2m}\Delta\psi = E\psi$ for the quantized billiard system. A detailed comparison of the measured resonances with the computed eigenvalues will be published elsewhere.

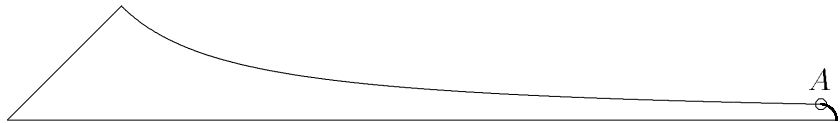

Figure 1: Shape of the truncated hyperbola billiard for $\alpha \approx 7.12$

The Green function for the Schrödinger equation with Dirichlet or Neumann boundary conditions can be approximated semiclassically ($\hbar \to 0$) by

$$G(\vec{q}'', \vec{q}', E) \approx \sum_{\text{cl. orbits}} \frac{1}{i\hbar\sqrt{2\pi i\hbar}} \sqrt{|D|}$$
$$\times \exp\left\{\frac{i}{\hbar} S(\vec{q}'', \vec{q}', E) - i\frac{\pi}{2}\mu\right\}. \tag{1}$$

This relation can be derived from the well-known Van Vleck formula by Fourier transformation [1]. From this one obtains the semiclassical contribution of classical orbits to the quantum mechanical eigenvalue spectrum by taking the trace ($\Omega$ denotes the billiard domain)

$$g(E) = \int_\Omega d^2q\, G(\vec{q}, \vec{q}, E), \tag{2}$$

whose imaginary part is proportional to the quantum mechanical level density

$$d(E) = \sum_{n=1}^{\infty} \delta(E - E_n) = -\frac{1}{\pi}\text{Im}\, g(E). \tag{3}$$

According to eq. (1), the Green function is approximated by a sum over all classical trajectories running from $\vec{q}'$ to $\vec{q}''$ with energy $E$. The phase of these oscillating terms is given by the classical action $S(\vec{q}'', \vec{q}', E)$, and an index $\mu$ counting twice the number of reflections on Dirichlet boundaries plus the number of conjugate points along the classical orbit. The amplitude is given by the square root of Van Vleck's determinant

$$D = -\det\begin{pmatrix} \partial^2 S/(\partial\vec{q}'\partial\vec{q}'') & \partial^2 S/(\partial E \partial\vec{q}'') \\ \partial^2 S/(\partial\vec{q}'\partial E) & \partial^2 S/\partial E^2 \end{pmatrix}, \tag{4}$$

which in two dimensions can be expressed by the upper right element of the classical $2 \times 2$ monodromy matrix $M$:

$$D = -\frac{1}{v'v''} \frac{\partial^2 S}{\partial q_2' \partial q_2''} = \frac{1}{v'v''} \frac{1}{M_{12}}. \tag{5}$$

Here $v'$ and $v''$ denote the particle's velocity in $\vec{q}'$ and $\vec{q}''$ respectively. The definition of the monodromy matrix and a method of computing it in case of billiard systems can be found in [6]. For billiard systems with twice continuously differentiable boundary, $M_{12}$ is a continuous function of energy and the points $\vec{q}'$ and $\vec{q}''$. The integration over configuration space in eq. (2) can therefore be approximated by focusing one's attention on the neighborhood of points $\vec{q} = \vec{q}' = \vec{q}''$ with a stationary phase, i.e.

$$\left[\frac{\partial S(\vec{q}'', \vec{q}', E)}{\partial \vec{q}''} + \frac{\partial S(\vec{q}'', \vec{q}', E)}{\partial \vec{q}'}\right]_{\vec{q}''=\vec{q}'=\vec{q}} = \vec{p}'' - \vec{p}' = 0. \tag{6}$$

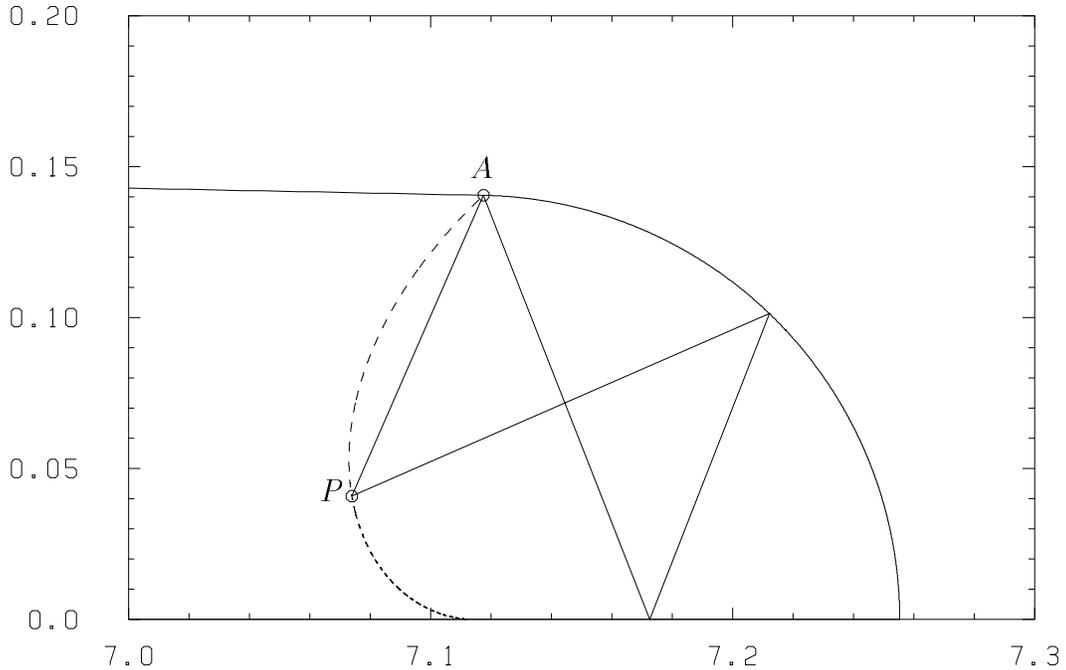

Figure 2: Example of a critical orbit with starting point $P$, and the critical curve (broken line) of its family

Thus the stationary points are the initial and final positions of trajectories which are closed in phase space - the periodic orbits. In their neighborhood, the slowly varying amplitude $\sqrt{|D|}$ can be regarded as constant and the action can be expanded up to second order in the coordinate variable $q_2$ which is perpendicular to the orbit itself:

$$S(\vec{q},\vec{q},E) = S + \frac{1}{2}\kappa q_2^2 + \cdots \qquad (7)$$

Following this program, one eventually obtains Gutzwiller's trace formula [1]

$$g_{\text{po}}(E) \approx \sum_{\text{po}} \frac{1}{i\hbar} \frac{L_0}{v} \frac{1}{\sqrt{|2 - \text{Tr}M|}} \exp\left\{\frac{i}{\hbar}S - i\frac{\pi}{2}\sigma\right\}, \qquad (8)$$

where $\sum_{\text{po}}$ is a sum over all periodic orbits with energy $E$, $L_0$ is the geometrical length of the primitive periodic orbit (po), and $\sigma = \mu - \frac{1}{2}(\text{sgn}(\kappa) - 1)$ is a coordinate invariant Morse index attached to the periodic orbit [8].

The boundary of the truncated hyperbola billiard contains a so-called *critical point* at $A = (\alpha, \frac{1}{\alpha})$, where the boundary curvature changes discontinuously. As the monodromy matrix of a classical trajectory depends on the boundary's curvature in the reflection points, the closed, but non-periodic orbits being reflected in such a critical point have to be examined thoroughly. We will call them *critical orbits* (co), see fig. 2. They are typically forming one-parameter families. This can be seen by continuously varying the angle of incidence at the critical reflection point $A$. The set of initial and final positions of such a family $F$ of closed critical orbits yields a smooth curve $C_F$ which we call *critical curve*, see fig. 2. In their vicinity the stationary phase method fails due to the discontinuous amplitude factor $\sqrt{|D|}$.

We now determine the contribution of critical orbits (strictly speaking: of orbits in their vicinity) to the trace $g(E)$. This can be done analogously to the derivation of Gutzwiller's trace formula indicated above. A curvilinear coordinate system $(s_1, s_2)$ is introduced along each critical curve, $s_1$ being the arc parameter and $s_2$ the coordinate perpendicular to the critical curve. Along the direction of $s_2$, two separate integrations have to be performed - for the orbits with the crucial reflection point to the left and to the right of the critical point $A$ respectively. In each $s_2$-integration, the matrix element $M_{12}$ is a continuous, slowly changing function of $s_2$ and can thus be approximated by the constant value $M_{12}^{\pm} = \lim_{s_2 \to 0} M_{12}$ with $\pm$ denoting the different values for the two families of orbits, depending on the different curvature radii $R^+$ and $R^-$ in the critical point.

Now the action is expanded up to second order in $s_2$

$$S(\vec{s}, \vec{s}, E) = S + \theta s_2 + \frac{1}{2}\kappa s_2^2 + \cdots \quad (9)$$

(note that $\theta \neq 0$ for non-periodic orbits), and the integration interval is extended to $[0, \infty)$. At this point, special care has to be taken in the case of non-periodic orbits. For $\text{sgn}(\theta) \neq \text{sgn}(\kappa)$ there is a stationary point of the approximated classical action at the point $s_2 = -\frac{\theta}{\kappa}$ inside the integration interval. But either the genuine action has no stationary point at all, or its contribution is already included in Gutzwiller's trace formula (8) for periodic orbits. In both cases the integral $\int_0^\infty ds_2$ has to be regularized by subtracting from it the contribution $\int_{-\infty}^{\infty} ds_2$ of the "false" stationary point.

Keeping this in mind, the $s_2$-integrations can be performed explicitly resulting in the sum of two error functions. The details of these calculations can be found in [9]. The final result is

$$\begin{aligned}g_{\text{co}}(E) &\approx \sum_F \int_{C_F} ds_1 \frac{1}{2i\hbar v} \sum_{\epsilon=\pm} \frac{1}{\sqrt{|\kappa^\epsilon M_{12}^\epsilon|}} \text{sgn}(\theta^\epsilon) \\ &\times \text{sgn}(\kappa^\epsilon) \exp\left(\frac{i}{\hbar}S - i\frac{\pi}{2}\sigma^\epsilon - iU^\epsilon\right) \\ &\times \text{erfc}\left(\sqrt{-iU^\epsilon}\right). \end{aligned} \quad (10)$$

In this trace formula, $\sum_F$ is a sum over all one-parameter families $F$ of non-periodic critical orbits. $\int_{C_F} ds_1$ is taken along each critical curve and cannot be evaluated analytically (in contrast to the periodic-orbit theory). It is therefore calculated numerically, see below. The sum $\sum_{\epsilon=\pm}$ refers to the two types of "almost critical" orbits with their crucial reflection point to the left ($\epsilon = +$) or to the right ($\epsilon = -$) of the critical point $A$ respectively. The Morse index $\sigma^\epsilon$ is again defined by $\sigma^\epsilon = \mu^\epsilon - \frac{1}{2}(\text{sgn}(\kappa^\epsilon) - 1)$ and $U^\epsilon$ is an abbreviation for $U^\epsilon = \frac{(\theta^\epsilon)^2}{2\hbar\kappa^\epsilon}$. Note that it is in general not allowed to replace the error function in eq. (10) by its asymptotic approximation since $U^\epsilon = 0$ is possible for certain orbits, e.g., for those orbits which start and end in a self-conjugate point, see also [9].

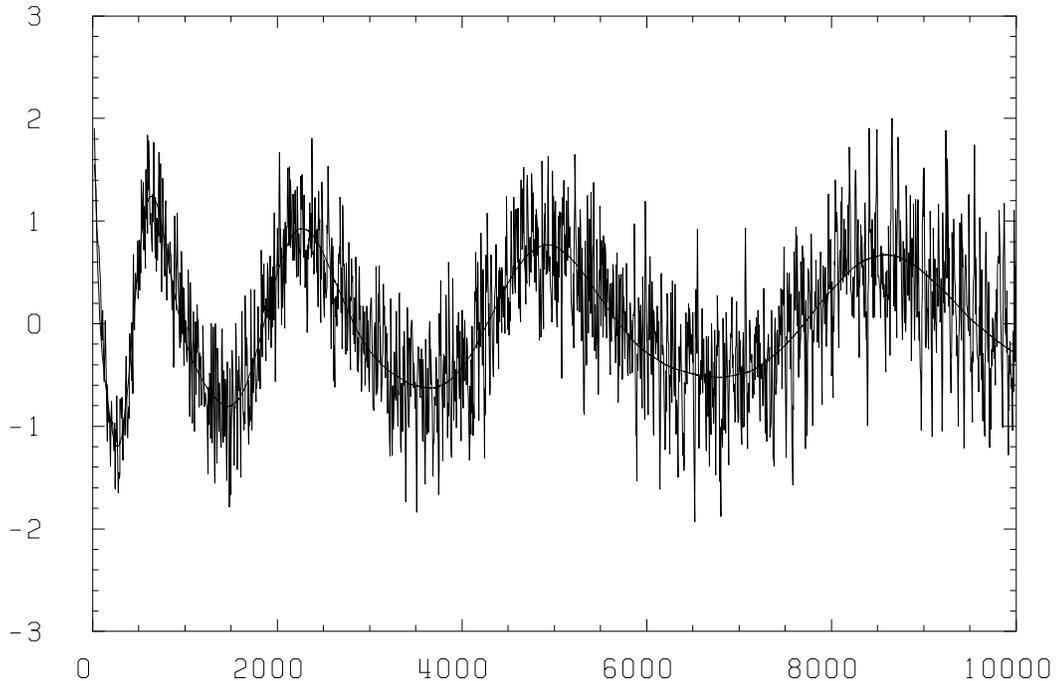

Figure 3: $N(E) - \frac{1}{2} - \overline{N}(E)$ and $N_{\text{co}}(E)$ for $E = E_n$

In order to test this formula numerically, the contribution $d_{\text{co}}(E)$ of non-periodic critical orbits to the level density has been calculated using eq. (3). Integrating this density gives the contribution $N_{\text{co}}(E)$ of critical orbits to the spectral staircase function $N(E) = \#\{n \,|\, E_n \leq E\}$. The integration constant is chosen such that $\lim_{E \to \infty} N_{\text{co}}(E) = 0$.

The energy spectrum of the truncated hyperbola billiard has been calculated between $E = 0$ and $E = 10\,000$ (in natural units $\hbar = 2m = 1$) using the boundary element method. In this interval all 1851 eigenvalues have been computed with high accuracy. No degeneracies have been found.

In fig. 3 the fluctuating part $N(E) - \frac{1}{2} - \overline{N}(E)$ is evaluated at the eigenvalues $E = E_n$, with

$$\overline{N}(E) = \frac{A}{4\pi} E - \frac{L}{4\pi} \sqrt{E} + C \qquad (11)$$

being the generalized Weyl's law for this billiard system. Here $A$ and $L$ denote the area and perimeter respectively of the truncated hyperbola billiard and $C = \frac{29}{96}$, see [9]. The unusual long-range modulation of the spectral fluctuation visible in this figure can be explained by the contribution $N_{\text{co}}(E)$ of critical orbits. The contribution of all 5 families of orbits with less than 5 reflections altogether has been calculated numerically. The total contribution is also presented in fig. 3, showing a very good agreement over the whole examined energy range. Subtracting the "correction term" $N_{\text{co}}(E)$, see fig. 4, yields spectral fluctuations typical for billiard systems and should semiclassically be explainable by Gutzwiller's periodic-orbit theory.

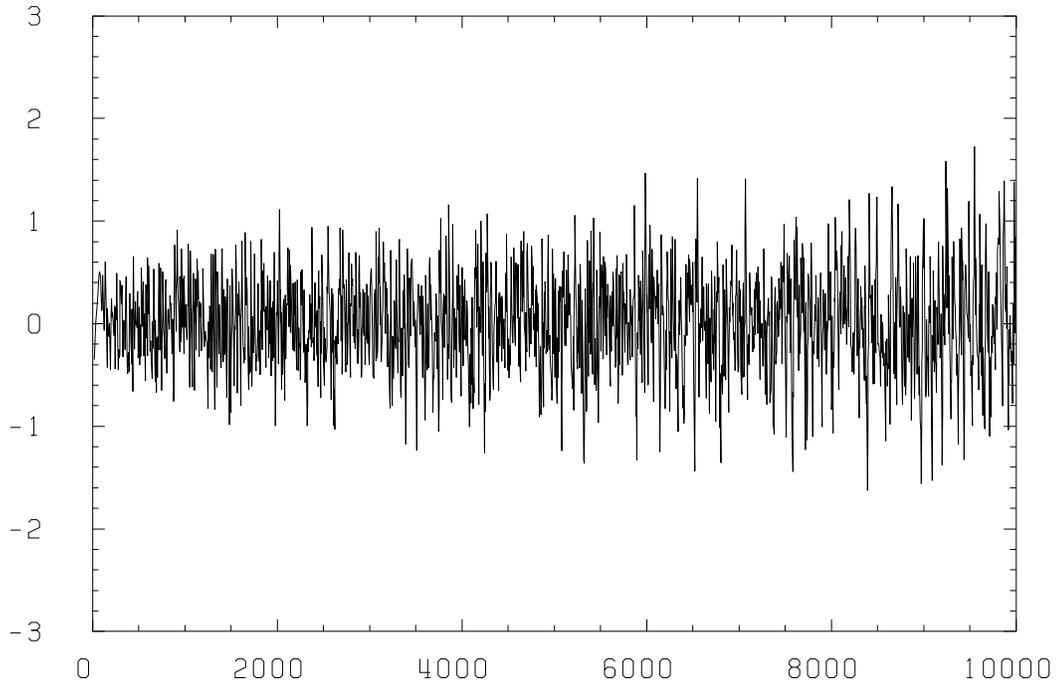

Figure 4: $N(E) - \frac{1}{2} - \overline{N}(E) - N_{\text{co}}(E)$ for $E = E_n$

The numerical results presented in this Letter clearly show that the contribution of non-periodic orbits to the integrated level density cannot be neglected for the truncated hyperbola billiard. The main reason for their numerical significance in the energy interval $0 < E < 10\,000$ is their geometrical length of approximately $\frac{2}{\alpha}$ compared to the shortest periodic orbit with a length of 2. It is therefore expected that their effect becomes even bigger for larger values of $\alpha$. We also predict an energy dependent critical orbit contribution to other quantities like the number variance, spectral rigidity, and the cosine-modulated heat kernel.